# Transmission IR Microscopy for the Quantitation of Biomolecular Mass In Live Cells

Yow-Ren Chang, Seong-Min Kim, and Young Jong Lee*

Biosystems and Biomaterials Division, National Institute of Standards and Technology, Gaithersburg, MD 20899, USA

**Abstract**: Absolute quantity imaging of biomolecules on a single cell level is critical for measurement assurance in biosciences and bioindustries. While infrared (IR) transmission microscopy is a powerful label-free imaging modality capable of chemical quantification, its applicability to hydrated biological samples remains challenging due to the strong water absorption. We overcome this challenge by applying a solvent absorption compensation (SAC) technique to a home-built quantum cascade laser IR microscope. SAC-IR microscopy improves the chemical sensitivity considerably by adjusting the incident light intensity to pre-compensate the IR absorption by water while retaining the full dynamic range. We demonstrate the label-free chemical imaging of key biomolecules of a cell, such as protein, fatty acid, and nucleic acid, with sub-cellular spatial resolution. By imaging live fibroblast cells over twelve hours, we monitor the mass change of the three molecular species of single cells at various phases, including cell division. While the current live-cell imaging demonstration involved three wavenumbers, more wavenumber images could measure more biomolecules in live cells with higher accuracy. As a label-free method to measure absolute quantities of various molecules in a cell, SAC-IR microscopy can potentially become a standard chemical characterization tool for live cells in biology, medicine, and biotechnology.

**Significance**: Biomolecular mass is fundamental information of a cell. This paper demonstrates the first transmission-mode IR images of live cells by a newly developed non-synchrotron IR microscope, enabling the measurement of per-cell mass of key biomolecules over time. Monitoring SI-traceable chemical quantities in live cells can become a standard critical method in cell-based drugs and therapies.

Characterizing and quantifying biomolecules in cells is critical for understanding cellular functions,[1] macromolecular crowding,[2,3] advanced biomanufacturing of therapeutics,[4] and disease progression and diagnostics.[5,6] However, the quantification of absolute mass in single live cells remains a measurement challenge. Traditional biochemical assays[7] and mass spectrometry-based analyses are extremely sensitive,[8] but their ensemble-based, destructive approach is unsuitable for characterizing highly heterogeneous, live cell systems. Optical imaging techniques such as fluorescence microscopy can discriminate individual cells with high molecular specificity to target markers. Still, the quantification of cell-consisting molecules is complicated by a lack of controllability in labeling efficiency and photobleaching.[9,10] Alternatively, spontaneous or coherent Raman imaging[11–13] can identify molecules without labeling by vibrational signatures; however, weak emission signals are system-dependent and, thus, yield only relative quantitation and require internal references of known concentrations for absolute quantification.[13]

Infrared (IR) absorption-based imaging can simultaneously identify label-free and quantify the absolute concentrations of molecules of a cell, including proteins, fatty acids, and nucleic acids, by detecting their spectral signatures with relatively high cross-sections.[14] Moreover, the measured value, absorbance, is system-independent, making the value SI (international system of units)-traceable and, thus, interlaboratory-comparable. However, a fundamental challenge in IR absorption measurements of biological molecules is the broad and strong IR absorption by water.[15] In particular, the water bending mode at 1650 cm$^{-1}$ overlaps with the amide I absorption peak, a characteristic peak for protein quantification and secondary structure analysis,[16] making IR unsuitable for quantifying this key biomolecule.

This challenge to IR imaging has been passively mitigated either by reducing the optical path length,[17,18] employing attenuated total reflectance (ATR) configurations,[19,20] or using a synchrotron radiation for increased light intensity.[21] However, thinner sampling chambers (<10 μm) suffer from physical compression of cells and difficulty in microfluidic control for live cell imaging applications. Despite the easy coupling of ATR detection with a live-cell chamber, ATR-IR can sense only a few micrometers near a substrate surface, representing only part of a cell. Bright synchrotron sources to enable transmission through water have been used for live cell IR imaging.[22–24] However, the limited accessibility hinders the broad applicability of IR imaging.

Recently, readily available bench-top external-cavity quantum cascade lasers (EC-QCLs), which emit monochromatic tunable light in mid- and far-IR (800 – 3000 cm$^{-1}$), have enabled new discrete frequency infrared microscope designs.[25,26] Notably, QCL-based IR absorption microscopy demonstrates high speed, aberration correction, and high signal-to-noise (SNR).[27,28] Other QCL-based IR microscopy methods based on indirect photothermal IR signals have been demonstrated.[29,30] For example, IR-pump visible-probe (refraction [29,31,32] and fluorescence [29,33,34]) approaches enabled non-contact IR absorption imaging with sub-micrometer spatial resolution and were used for live cell imaging. However, the signal changes caused by a photothermal effect are sample- and system-dependent and, thus, are not readily convertible to absolute concentrations or mass.

Despite all these advances, the broad and strong IR absorption by water near the amide I band still limits the dynamic range of the detection systems and obscures the absorption signal of biomolecules of interest. To address this critical limitation, we recently demonstrated that a simple optical technique called solvent absorption compensation (SAC) could eliminate water absorption contribution optically and enhance SNR by >100 times for the amide I absorption peak compared to a conventional IR transmission measurement.[35] In the present work, we introduce SAC into a home-built sample-scanning, transmission mode QCL-IR microscope by attenuating the intensity of the mid-IR beam via a rotating polarizer as a function of wavelength. SAC pre-compensates for solvent absorption and ensures a constant transmission intensity across the entire wavenumber range spanning from 900 cm$^{-1}$ to 1776 cm$^{-1}$. This wide range enables label-free imaging of multiple biomolecules, including protein, fatty acid, and nucleic acid, inside a single cell. We compare protein mass per cell between fixed and live cells using the amide I absorption peak of proteins near 1650 cm$^{-1}$, which was not available by conventional non-SAC-IR approaches. We characterize system performance and demonstrate that a simple, direct transmission-based IR measurement can be utilized for live, intercellular absolute protein mass measurements.

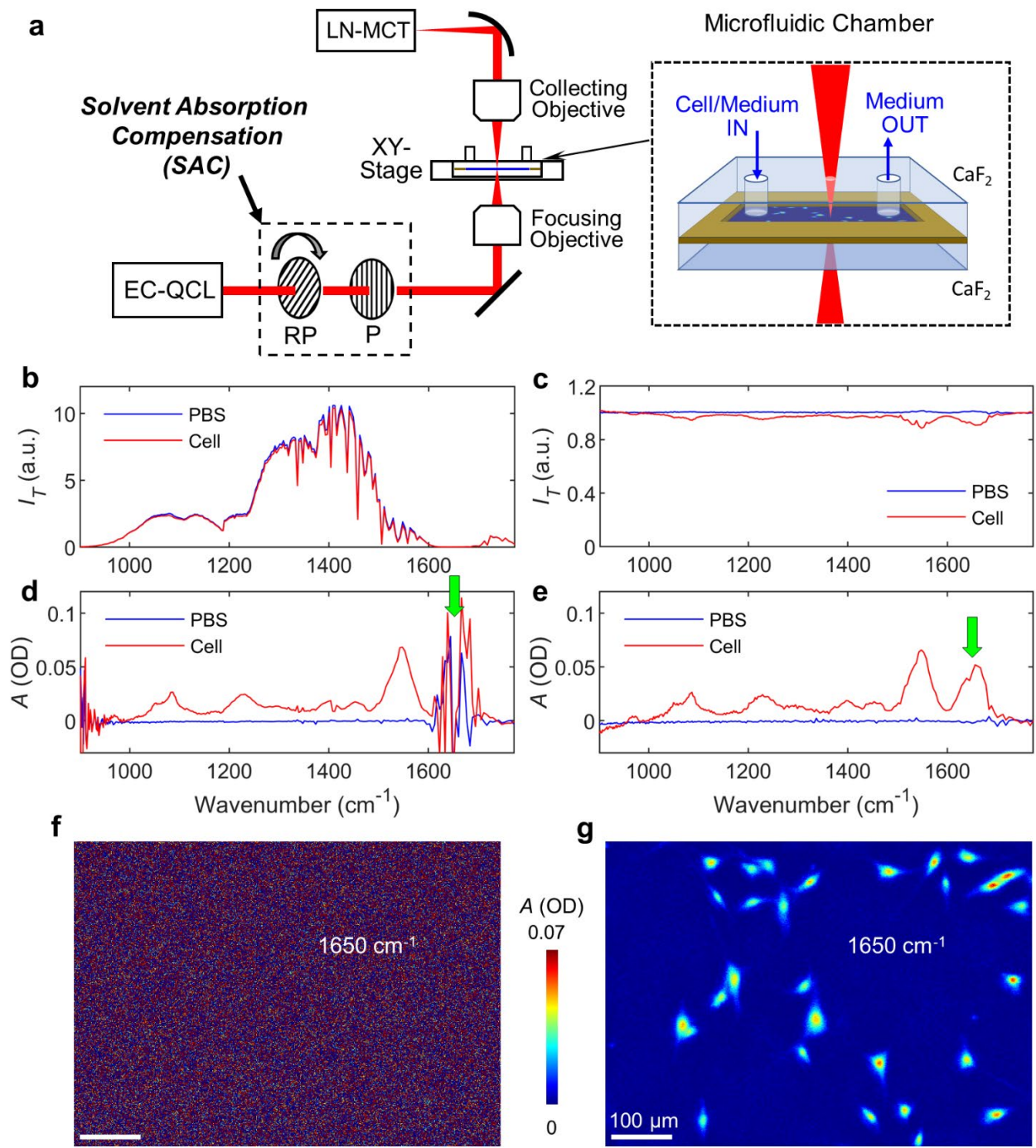

**Figure 1**. (**a**) Schematic of solvent absorption compensation infrared (SAC-IR) microscopy setup. The laser intensity is controlled as a function of wavelength by a pair of rotating and fixed polarizers. EC-QCL: external-cavity quantum cascade laser; RP:

rotating polarizer; P: fixed polarizer; and LN-MCT: liquid nitrogen-cooled mercury-cadmium-telluride detector. The inset illustrates a microfluidic sample chamber consisting of two 1-mm thick $CaF_2$ windows and a 25 µm spacer. One of the $CaF_2$ windows has drilled holes for introducing cells and media. (**b**,**c**) Spectra of transmitted light intensity ($I_T$) in a fixed fibroblast cell region (red) and a phosphate buffer saline (PBS) region (blue) without SAC (**b**) and with SAC (**c**). (**d**,**e**) Absorbance spectra of a PBS (reference, blue) and a fixed fibroblast cell (red) without SAC (**d**) and with SAC (**e**). (**f**,**g**) Absorbance images of fixed fibroblast cells, measured at 1650 cm$^{-1}$ without SAC (**f**) and with SAC (**g**).

Figure 1a shows the schematic of the SAC-IR microscope (SI for detailed description). Briefly, for SAC, the IR beam passed through an IR polarizer (ISP Optics) mounted on a rotation stage (Newport) and then back through a fixed IR polarizer; rotating the first polarizer modulated the incident light intensity as a function of wavelength. Sample chambers used in this work consisted of a 1 mm thick $CaF_2$ slide (Crystran), a 25 µm spacer tape (3M), and a 1 mm thick $CaF_2$ slide with drilled holes for cells and media introduction. For SAC, the transmitted beam was moved to a fluid-only blank region (termed reference region), and for each wavenumber, the rotating polarizer angle was set to achieve a constant setpoint intensity over the entire wavenumber scanning range. (see Figure S1).

To demonstrate the need for SAC, we compared the transmitted light intensity $I_T$ of a reference region (Figure 1b, the blue line) to the transmission intensity from an image pixel near the center of a fixed NIH 3T3 fibroblast cell (Figure 1b, the red line, see SI for cell culture details). Figure 1b shows that without SAC, water strongly absorbs IR in the region of 1600 cm$^{-1}$ – 1700 cm$^{-1}$, and the intensity difference between a reference (PBS-only region) and sample (PBS and fibroblast cell) is minimal. In comparison, Figure 1c demonstrates the SAC implementation; the reference intensity spectrum (PBS, the blue line) is near constant, and there is a resolvable intensity difference between the reference and sample. We also note that IR below 1000 cm$^{-1}$ is weak due to strong absorption by the $CaF_2$ windows, but SAC compensates for substrate IR absorption, too.

Figures 1d,e show absorption spectra in a reference region (blue) and near a cell center (red). The SAC-implemented IR spectra of the cell in the 1600 cm$^{-1}$ –1700 cm$^{-1}$ region are observable with a high SNR. We then constructed absorption images of hydrated fixed fibroblast cells (see Figure S2 for details). Absorption images (Figures 1f,g) show that cells are unresolved from the background because the strong absorption of water overlaps with absorption from biomolecules within the fibroblast cell. In SAC (Figure 1g), water absorption is compensated, and the absorption by analyte molecules in the cells becomes resolvable from the background. Figure S3 shows histograms of absorbance values; the broad distribution of the background in non-SAC overwhelms the signal of cells, whereas, in SAC, the cells' absorbance values can be readily resolved from the narrowed main distribution. We estimate the pixel-to-pixel (spatial) noise at 1645 cm$^{-1}$ in the background region (Figure S3) of non-SAC absorption images to be 25 mOD. In comparison, the spatial noise in SAC images is 2 mOD in, a >10-fold improvement compared to non-SAC. Similarly, the spectral noises measured in the 1623 cm$^{-1}$ – 1672 cm$^{-1}$ region are 20 mOD and 2 mOD for non-SAC and SAC, respectively (Figure S4).

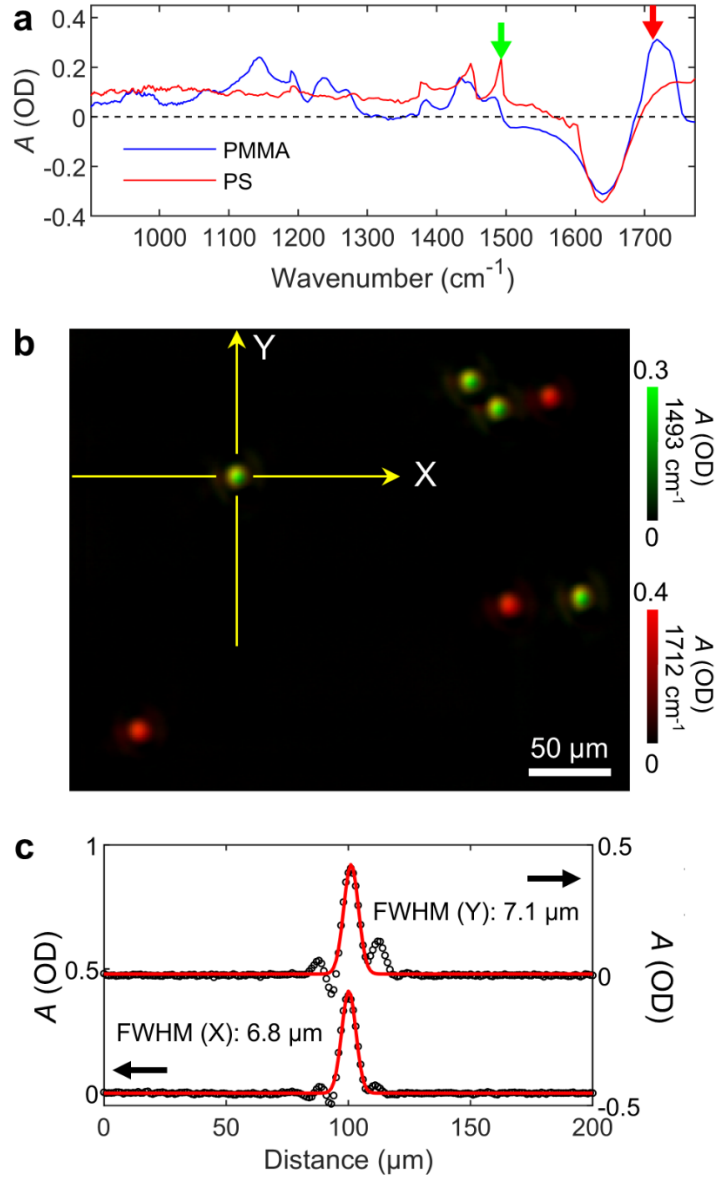

**Figure 2.** System performance using a mixture of 5-μm diameter polystyrene (PS) and poly(methyl methacrylate) (PMMA) microparticles in water with a 25 μm spacer. (**a**) IR spectra of the PS and PMMA microparticles using SAC. (**b**) Composite image with absorbances at 1493 cm$^{-1}$ (green) and 1712 cm$^{-1}$ (red) using SAC. (**c**) Line scans of the indicated PS particle in panel **b**.

Figure 2 demonstrates system performance in space and frequency using a mixture of polystyrene (PS, Phosphorex) and poly(methyl methacrylate) (PMMA, Phosphorex) particles dispersed in water. The nominal diameters of both particles are 5 μm. Similar to Figure 1, the absorbance of each image pixel was calculated by the reference (water) region. Thus, the apparent IR absorbance spectra of the two types of microparticles are shown in Figure 2a. Both spectra show a strong negative absorbance region centered near 1645 cm$^{-1}$, indicating that the transmission is higher at a pixel than the background region. This apparent negative absorbance occurs when the matter in the beam path absorbs less light of the specific wavenumber than the replaced water.[36,37] In addition to the water exclusion effect, the Mie scattering makes it difficult to quantitatively separate the inherent vibrational absorption spectrum and the scattering from an observed transmission spectrum. Even with the complex spectral mixing, the strong absorption peak centered near 1712 cm$^{-1}$ observed from PMMA particles can be assigned to the C=O stretching mode from the easter group of PMMA.[38] Similar to PMMA particles, the apparent IR absorption spectrum of PS particles in Figure 2a shows the strong negative dip at 1645 cm$^{-1}$ due to the water exclusion effect and the positive baseline drift due to scattering. However, still, the signature peaks at 1493 cm$^{-1}$ and 1453 cm$^{-1}$ of the benzene ring of PS were observable.[39]

  Figure 2b shows a pseudo-labeled composite image of a microparticle mixture constructed with two absorbance images acquired at 1493 cm$^{-1}$ (green) and 1712 cm$^{-1}$ (red), corresponding to PS and PMMA, respectively. All particles in the composite image were readily identifiable as either PS or PMMA. We measured the full-width-half-maximum (FWHM) from line scans of multiple particles. The mean FWHMs from particles are 7.3 ± 0.2 μm in the fast-scanning direction (X, horizontal) and 7.0 ± 0.4 μm in the slow-scanning direction (Y, vertical), where the uncertainties indicate the standard deviation. This resolving power is close to the diffraction-limited resolution (0.5*λ*/*NA*) at 1650 cm$^{-1}$ with a 0.4 NA reflective objective.

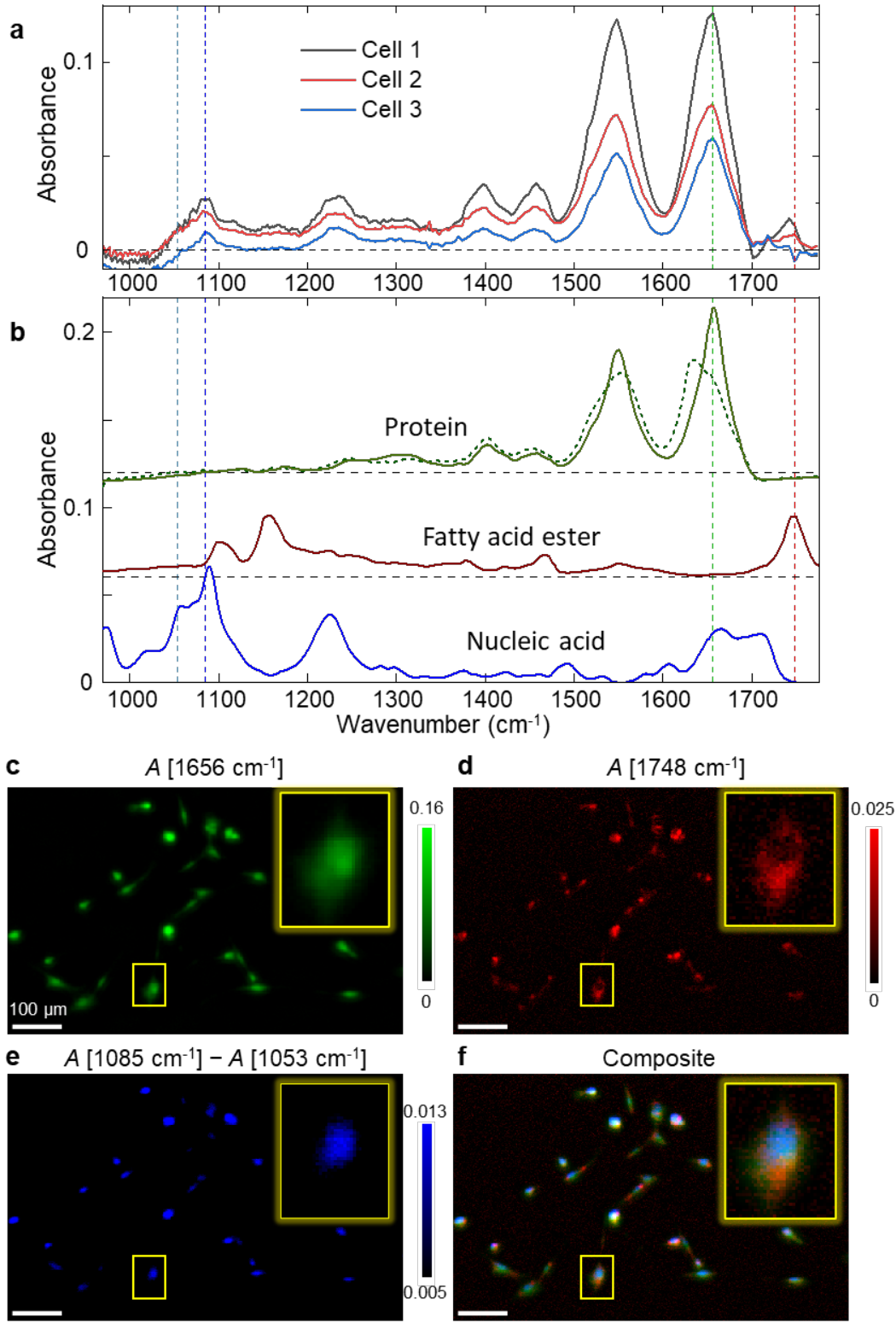

**Figure 3**. (**a**) SAC-IR spectra of three different fixed fibroblast cells. (**b**) IR spectra of BSA (solid green, rich in α-helix) and β-lactoglobulin (dotted green, rich in β-sheet) for protein, and herring DNA in water (blue) for nucleic acid. The proteins and DNA solutions were prepared in water. The spectrum of glycerol trioctanoate in $CCl_4$ and $CS_2$ (red) for fatty acid ester was downloaded from the NIST Chemistry WebBook (webbook.nist.gov). All absorption spectra were scaled to the concentration of 10 mg/mL. (**c–e**) Absorbance images at wavenumbers representative of (**b**) protein at 1656 cm$^{-1}$, (**c**) fatty acid at 1748 cm$^{-1}$, and (**d**) nucleic acid at 1085 cm$^{-1}$ and 1053 cm$^{-1}$. (**e**) Composite image of panels **c–e**. The arrows indicate the wavenumbers used to construct the images of panels **c–e**.

Next, we demonstrate label-free chemical imaging of key biomolecules in fixed cells. Figure 3a shows the absorbance spectra of three fixed fibroblast cells. For each spectrum of a cell, the sample stage was X-scanned at a fixed Y location, and it was repeated while the wavenumber changed. The line-averaged spectra of three different cells show similarity in IR peak position and width, although peak heights vary among cells. In Figure 3b, IR spectra were presented for comparison from representative cellular molecules, including protein, nucleic acid, and fatty acid ester. The strong peaks at 1650 cm$^{-1}$ and 1550 cm$^{-1}$ from the cells show a strong correlation with the amide I and II peaks of protein, respectively. On the other hand, the peaks observed at 1085 cm$^{-1}$ and 1230 cm$^{-1}$ correspond to the symmetric and antisymmetric $PO_2^-$ stretching modes of nucleic acid, respectively.[40] The weak peak at 1745 cm$^{-1}$ can be attributed to the ester group in phospholipid or fatty acid ester. Although various types of ester molecules exist in cells, we will call the 1745 cm$^{-1}$ peak for fatty acid, for simplicity, in this paper. Figures 3c–e show absorbance images constructed at the most relevant wavenumbers of protein, fatty acid, and nucleic acid, respectively. Because of the relatively weak absorbance of nucleic acid compared to the baseline, Figure 3e for nucleic acid was constructed by absorbance difference between 1085 cm$^{-1}$ and 1053 cm$^{-1}$. The composite image in Figure 3f shows the spatial distribution of the three biomolecules. The blue-colored feature is located at the center of cells, while protein is distributed over the entire cell. On the other hand, fatty acid is localized in non-nuclear regions.

Among the IR peaks that represent biomolecules in the cell spectra, we focused on the absorbance peak at 1650 cm$^{-1}$, which is dominantly due to the amide I peak of protein. Also, the absorption cross sections of major biomolecules, such as nucleic acid and fatty acid, are lower than that of protein at 1650 cm$^{-1}$, as shown in Figure 3b. Some carbohydrates contain peptide bonds, e.g., sialic acid, but their concentration in a cell is much lower than proteins. Thus, for simplicity, we assumed the absorbance at 1650 cm$^{-1}$ originated from proteins and used it to determine the absolute protein mass per cell.

If cell height is known, measured absorbance will lead to the concentration of molecules in a cell. However, the current imaging system could not measure absorbance and height simultaneously during live cell imaging. Instead, we determine the mass per cell ($m_c$) from the absorbance sum ($\Sigma A$) and the absorption coefficient ($\varepsilon$) with the following relation,

$$m_\text{c} = c \times V = (\bar{A}/\varepsilon \bar{l}) \times (\bar{l}S) = \frac{1}{\varepsilon}\bar{A} \times S$$

(1)

where $c$ is the concentration, $V$ is the cell volume, $\varepsilon$ is the absorption coefficient, $\bar{l}$ is the mean cell height, and $\bar{A}$ is the mean absorbance over the cell area, $S$. A cell area was determined by segmenting cells in absorption images with the threshold of 99% of the cumulative distribution of the background absorbance values (see SI for details).

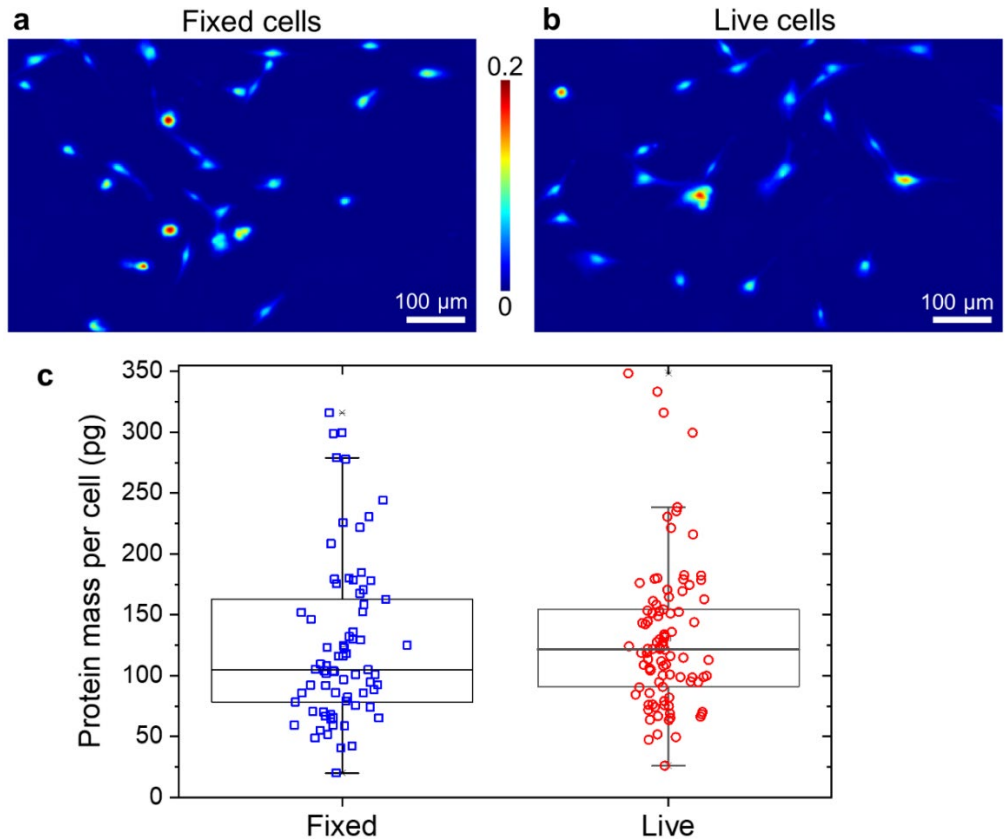

**Figure 4.** (**a,b**) Comparison of SAC-IR images of fixed and live fibroblast cells. Both images were acquired at the protein amide I peak at 1650 cm$^{-1}$. (**c**) Scatter plots of protein mass per cell for 74 fixed and 94 live cells. The median (mean) values are 105 pg (126 pg) and 122 pg (131 pg) for fixed and live cells, respectively. The boxes indicate the 25% and 75%, and the horizontal lines indicate the median values.

To calculate the total protein mass per cell, we used the absorption coefficient, $\varepsilon$ = 0.28 μm$^2$ pg$^{-1}$ at 1650 cm$^{-1}$, calculated from the spectra of two non-glycosylated proteins with different secondary structures, shown in Figure 3b. Their mean $\varepsilon$ value at 1650 cm$^{-1}$ can be considered to represent the mixture of proteins with various secondary structures. Using Eq. (1), we calculated the total protein mass per cell for both fixed and live cells. Figure 4c shows that the mass spans from 20 pg to 350 pg, and the distributions seem non-Gaussian. The median value of the fixed cells' protein mass per cell is 105 pg from 74 cells. This is slightly lower than that of live cells' value (122 pg from 94 cells). However, considering the breadth of the distributions, it is difficult to conclude whether the protein mass is reduced by fixation or not. As a sanity test, we compared the protein mass per cell by varying the threshold level for cell segmenting. Figure S5 shows that the mean protein mass per cell is reduced by 6% when the threshold is elevated from 1 σ to 3 σ, where σ is the standard deviation of the background absorbance distribution. The range of protein mass per cell is consistent with previously reported values from whole-cell protein mass measurements,[41] UV absorption microscopy,[42,43] and stimulated Raman scattering (SRS).[13] Thus, despite the many assumptions and simplifications, the absorbance at 1650 cm$^{-1}$ can be useful for the quantification of protein mass in live cells.

Next, we used SAC-IR microscopy to image live cells every 12 min for 12 h at three representative wavenumbers (1656 cm$^{-1}$, 1745 cm$^{-1}$, and 1230 cm$^{-1}$) for protein, fatty acid ester, and nucleic acid, respectively. For nucleic acid, we used absorbance at a single wavenumber of 1230 cm$^{-1}$ instead of absorbance difference, as shown in Figure 3, to increase the imaging frame rate and take advantage of slightly higher absorbance signals. The absorbance images sequentially acquired for the three wavenumbers were combined into a composite image every 12 min. The time-lapse of entire IR absorbance composite images for 12 h can be found in SI Movie 1. Most cells were agile, and some underwent cell division. Figure 5a shows IR images at three different times among the total 62 frames. We show trajectories of three isolated and in-frame,cells during the entire imaging experiment n in Figures 5b–d. The cell labeled as **b** showed cell division. The cell **c** was about to finish cell division in the last frames. On the other hand, cell **d** did not show division but showed fluctuation in size while migrating. We monitored the fluctuation of the absorbances at the three wavenumbers corresponding to protein, fatty acid, and nucleic acid. Figures 5e–g show the total absorbance per cell ($\bar{A} \times S$) as a function of time for the three wavenumbers. From Figures 5e,d, the absorbance tracks of the three biomolecules show a strong correlation, in particular, three to five hours before cell division, when all three absorbances began to increase, likely to prepare cell division. On the other hand, the non-dividing cell **d** shows little correlation between the three absorbance profiles. For proteins, the total absorbance was converted into the absolute mass per cell using Eq. (1) and labeled on the right. Unlike absorbance at 1656 cm$^{-1}$ for proteins, single-frequency absorbances at 1745 cm$^{-1}$ and 1230 cm$^{-1}$ can be contributed by other chemicals than fatty acid and nucleic acid. Thus, we did not convert the total absorbances at the two frequencies into the total masses of fatty acid and nucleic acid. However, if sufficiently more spectral images are available, their absolute per cell masses could be monitored as well. An IR transmission imaging system with a faster imaging speed and more frequencies could provide more quantitative, time-dependent information on the absolute masses per cell for various biomolecules in live cells.

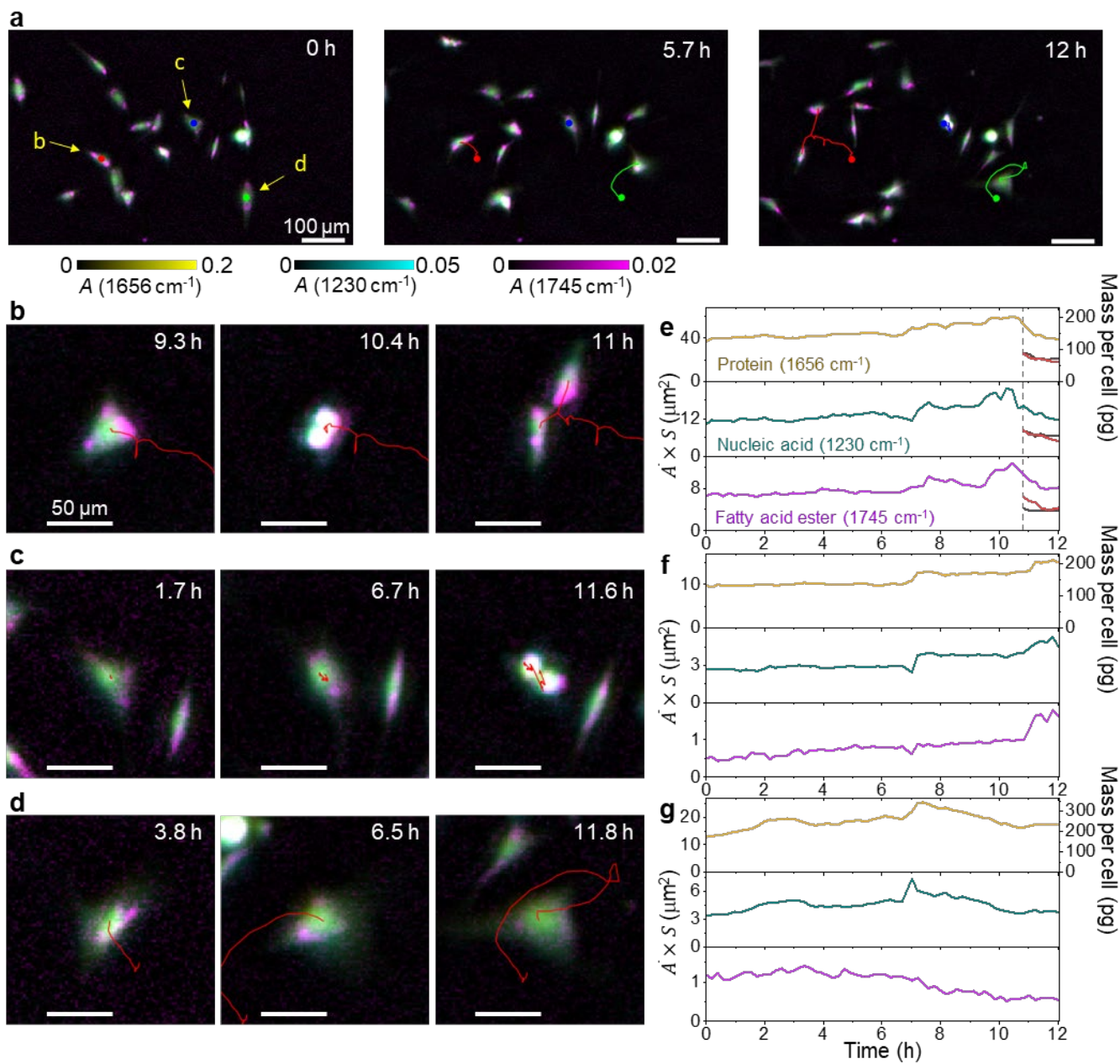

**Figure 5.** (**a**) Live fibroblast SAC-IR images acquired at three wavenumbers every twelve minutes for twelve hours. Yellow color represents absorbance associated with protein at 1656 cm$^{-1}$; magenta for fatty acid at 1745 cm$^{-1}$; and cyan for nucleic acid at 1230 cm$^{-1}$. (**b**–**d**) Enlarged image of the dividing cells (**b** and **c**) and the non-dividing cell (**d**), indicated in panel **a**. The centers of mass absorbance of the marked cells are plotted as time traces. (**e**–**g**) Tracking total absorbance per cell ($\bar{A} \times S$) of the cells in panel (**b**–**d**), respectively, measured at the three different wavenumbers. In (**e**), the red and black solid lines after 10.8 h are the $\bar{A} \times S$ values of the daughter cells.

In summary, we have demonstrated the non-synchrotron IR transmission imaging of live cells using the home-made SAC-IR microscopy technique. The SAC approach successfully mitigated the strong water absorption near the amide bands. Based on SAC-IR images measured at multiple frequencies, we successfully measured the quantity and distribution of three key biomolecules (protein, fatty acid, and nucleic acid) in fixed and live cells. In particular, using the total amide I band absorbance per cell, we measured the absolute per-cell mass of protein in multiple live fibroblast cells. By imaging live fibroblast cells over twelve hours, we monitor the per-cell mass change of the three molecular species during various phases, including cell division. The current demonstration was based on only three wavenumbers; an increase in imaging speed can yield additional wavenumber images while maintaining biologically relevant framerates, thus increasing the wealth of biomolecular measurements. SAC-IR transmission microscopy has the potential to become a standard characterization method of quantifying biomolecules with SI-traceability in live cells and hydrated tissues in biology, medicine, and biotechnology.

**Acknowledgments**

We thank Charles Camp for helpful discussions and Joy Dunkers for assistance with confocal microscopy. We thank the NIST Biomanufacturing Program and the National Research Council for financial support.

**Notes**

The authors declare no competing financial interest.

**Supporting Information**

# Transmission IR Microscopy for the Quantitation of Biomolecular Mass In Live Cells

Yow-Ren Chang, Seong-Min Kim, and Young Jong Lee*

Biosystems and Biomaterials Division, National Institute of Standards and Technology, Gaithersburg, MD 20899, USA

*SAC-IR microscope*

An external cavity QCL (EC-QCL, MIRCat, Daylight Solutions) equipped with a single QCL chip tunable from 900 cm$^{-1}$ – 1774 cm$^{-1}$ was used as an IR source. The laser was pulsed at 100 kHz with a 5% duty cycle. An IR polarizer (BSP-POL-4-11, ISP Optics) mounted on a rotation stage and a fixed IR polarizer was used to modulate the laser intensity (US patent US10184835B2). The beam was then focused onto the sample by a 25x, 0.4 NA reflective objective (LMM25X-UVV, Thorlabs). The transmitted light through the sample was collected by a 25x, 0.4 NA refractive objective (892-0004, Pike Technologies) and then focused onto a liquid nitrogen-cooled MCT detector (IR Associates) by an off-axis parabolic mirror. The detector signal was demodulated through a lock-in amplifier (LIA, SR865A, Stanford Research Systems) with a time constant of 100 µs. The LIA output signal was digitized with a DAQ device (BNC-2110, National Instruments). A sample chamber was placed in a temperature control unit (Ibidi) on a mechanical stage (MS-2000, ASI Scientific), which raster-moved the sample. For a fast X-scan, a TTL signal was sent at a designated X starting point by the stage controller to trigger data collection so that signals were used for imaging only from a constant velocity region. The linear stage speed was set at 5 mm/s for all imaging. PS/PMMA particle images were binned to 1 µm pixel sizes while cell images were binned to 2 µm pixel sizes. A typical 500 x 500 pixel image at a single wavenumber takes ~5 min. Hyperspectral IR images were collected by acquiring an XY image at a fixed wavenumber and repeating it at the next wavenumber step. Single-cell spectra were measured by repeating X-scans while the wavenumber was scanned with a step size was 2 cm$^{-1}$.

*Samples*

One-mm thick calcium fluoride (CaF$_2$, Crystran) windows were used as the "bottom" substrates for all samples. A 25 µm thick transfer tape (9969, 3M) was used as a spacer. For microparticle imaging, a chamber was sealed with a 0.2 mm thick CaF2 cover glass (Crystran). For cell imaging, a chamber was capped with a 1 mm thick CaF$_2$ window with two drilled holes, and nanoports (Idex) were used to attach 1/16" OD (outer diameter) tubing for cells and media introduction.

For microparticle samples, 5 µm diameter polystyrene (PS, Phosphorex) and 5 µm diameter poly(methyl methacrylate) (PMMA, Phosphorex) microspheres were mixed and dispersed in deionized water and introduced into a $CaF_2$ sample chamber.

For cell imaging, the $CaF_2$ substrates were prepared by immersing them in 70% ethanol for one hour and washed with PBS before being assembled into a flow chamber. NIH-3T3 fibroblasts (ATCC CRL-1658) were grown in DMEM supplemented with 10% serum (NBCS, Thermo Fisher) and seeded into the flow chamber. The flow chamber was placed in an incubator overnight to attach cells to the substrate. For fixed cell imaging, cells in the flow chamber were first washed with a PBS solution and then immersed in a fixing solution (Cytofix, BD) for 15 min, followed by washing with a PBS solution. The chamber was plugged before imaging.

For live cell imaging, the flow chamber with live cells was placed on a heated stage in the microscope with the temperature set to 36 °C. Every six hours, the old medium in the flow chamber was replaced with a fresh $CO_2$-saturated medium prepared in an incubator.

*IR imaging, image pre-processing, and absorbance calculation*

The QCL intensity passing through a water-filled 25 µm thick sample chamber was first adjusted by rotating the polarizer to reach a setpoint intensity across a wavenumber range (900 $cm^{-1}$ to 1774 $cm^{-1}$). Figure S1 shows a typical intensity profile with SAC. Raw intensity images, $I_{sample}$, are first normalized for any intensity drift during image acquisition (Fig. S2a-c). A finer normalization was performed by first masking off absorbing regions, fitting the remaining baseline to a third-order polynomial, and subtracting out any baseline variations (Fig. S2d-f) in both the horizontal and vertical directions of an image. We note that selecting an incorrect image mask can introduce image artifacts, such as shadowing around strongly absorbing objects, and that a visual inspection of the image processing is required. Finally, the absorbance is calculated line-by-line from the normalized intensity image. A final intensity mask is used to determine a blank region, then averaged, and is used as the reference intensity, $I_{ref}$. The absorbance, $A$, was then calculated as $A = \log_{10}(I_{ref}/I_{sample})$.

*Image segmentation*

Cells in absorbance images were segmented, and the sum absorbance per cell was converted to the total protein mass on a per-cell basis. The absorbance threshold level required for cell image segmentation was selected by inspecting the distribution of absorbance values in a blank region. The empty area was selected based on the intensity masks used in image pre-processing. Figure S6a shows the background absorbance, and we expected a Gaussian distribution centered near 0 with some distribution width representing the noise level in our system. Since the selection of a threshold level can be arbitrary, we evaluated the image segmentation quality and resulting mass measurement of several different metrics of the background distribution: the standard deviation, 95%, 99%, and 99.9%. Figure S6c shows that the four different threshold levels resulted in only 10 pg variation in the average value of the mass-per-cell distribution. For all mass measurements, we selected an absorbance threshold level that equated to 99% of the cumulative distribution of the background absorbance values. This value was approximately 0.006. After thresholding, a watershed transform was performed to segment neighboring cells (Fig. S6b). Cells that were too small or were cut off in the image were excluded from the analysis.

**Notes**

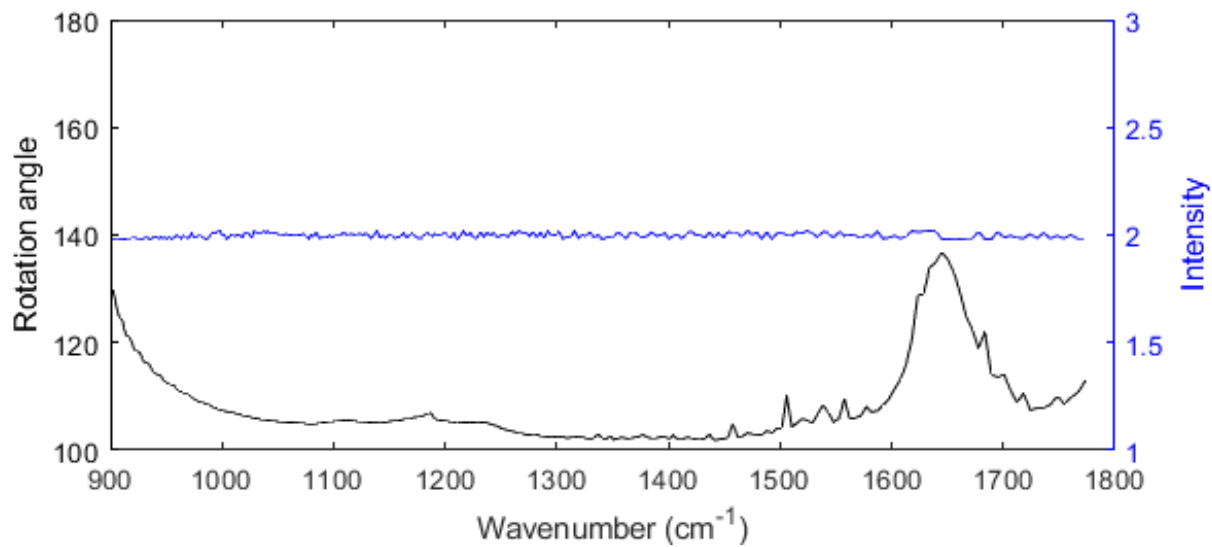

**Figure S1**. Example of polarization rotation calibration of a $CaF_2$ sample chamber filled with water with a 25 µm spacer. The left axis (black) shows a rotation angle of the moving polarizer set to reach a setpoint LIA signal (blue, right axis).

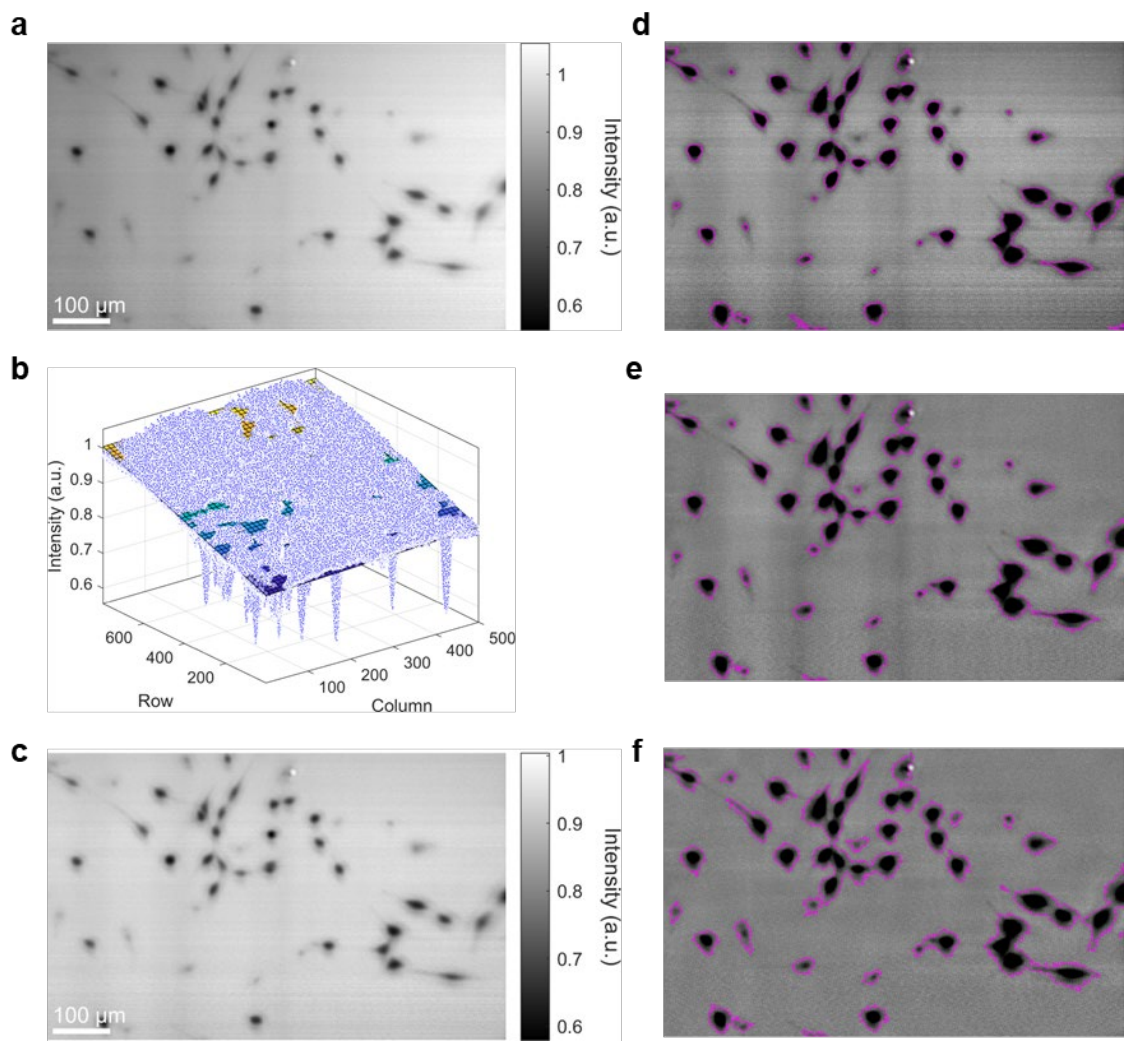

**Figure S2**. IR intensity image pre-processing steps. (a) The raw transmission IR image may have background tilt and line-by-line variations in intensity. A background plane correction is applied by fitting the entire intensity image to a flat plane (b), resulting in a background normalized intensity image (c). IR absorbing objects are masked (d), and a 3rd order polynomial is fitted to each horizontal line for baseline correction. This step is then repeated for the vertical direction (e). (f) A final intensity mask is applied for selecting a reference region.

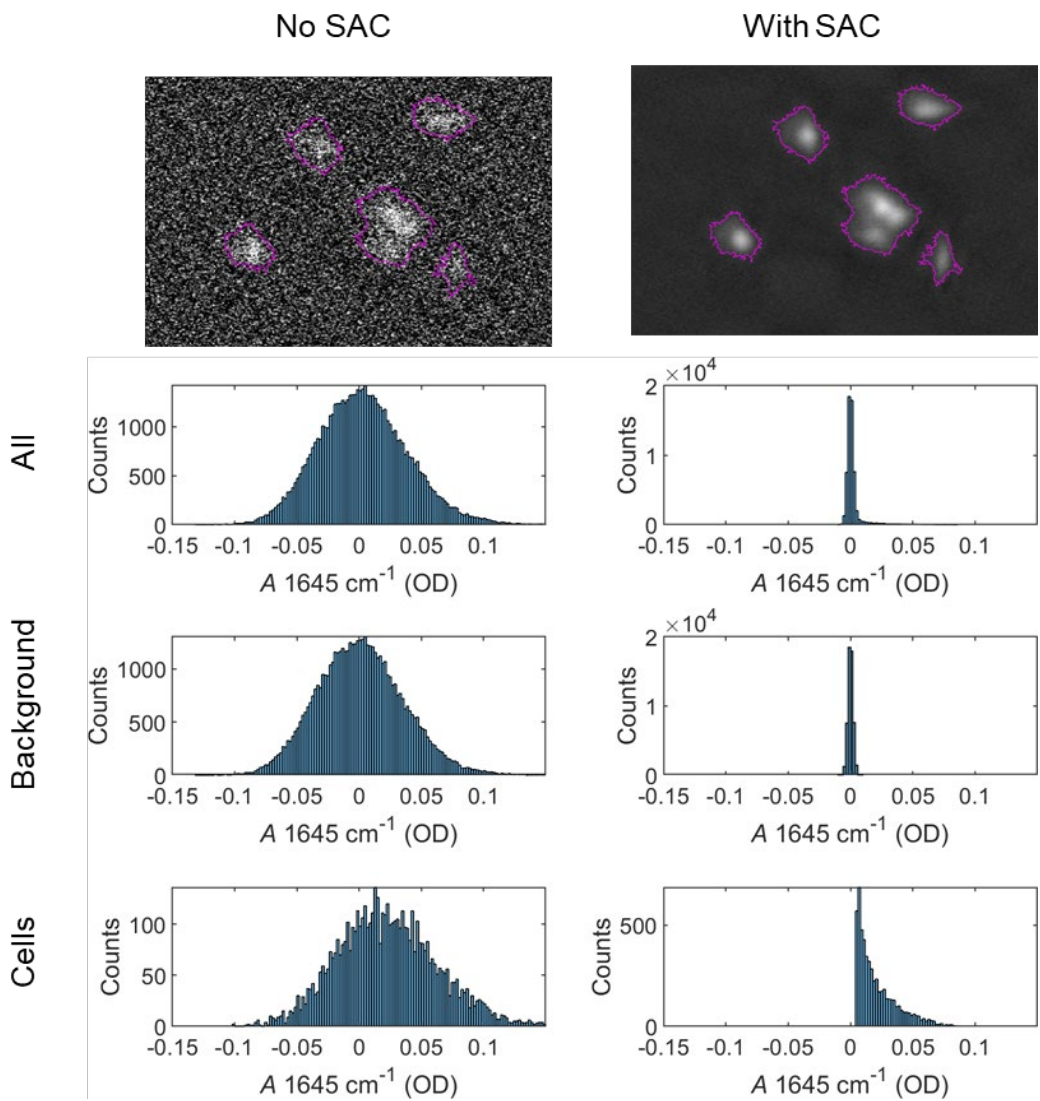

**Figure S3**. Histogram of 1645 cm$^{-1}$ absorbance values in IR absorbance images in Fig. 1d,e. An intensity threshold was used to segment absorbance images into the background and cells. The same masking areas were used for SAC and non-SAC images. The standard deviation of the background pixels was 25 mOD and 2 mOD for non-SAC and SAC images. The standard deviation of 2 mOD corresponds to the limit of detection of 0.8 mg/mL for BSA at the amide I peak.

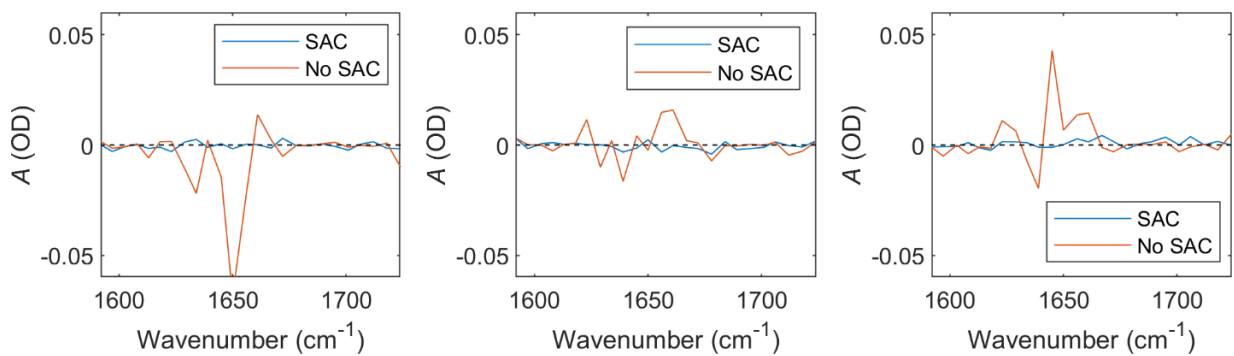

**Figure S4**. Example absorbance spectra of three single-pixel water regions with and without SAC. The dotted line is a guide to the eye and shows zero absorbance. The spectral noise was taken as the standard deviation of absorbance values in the 1623 cm$^{-1}$ – 1672 cm$^{-1}$ region. The mean spectral noise in the background region using the image mask in Figure S2 is 20 mOD and 2 mOD for non-SAC and SAC.

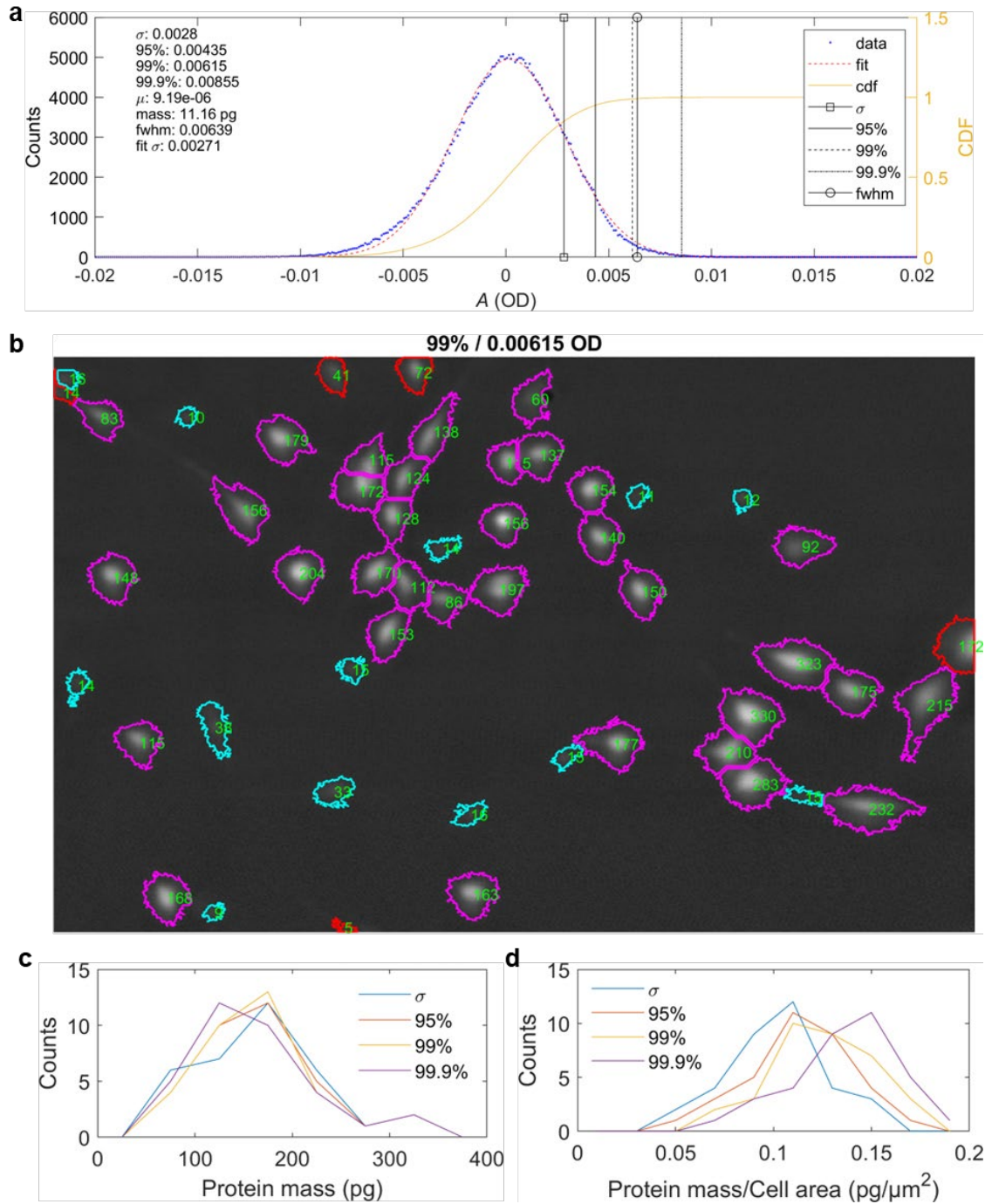

**Figure S5**. Selection of absorbance threshold level and sensitivity of threshold levels on protein mass measurements. (a) Histogram of background absorbance values from the unmasked area. The blue dots represent data; the dashed red line is a Gaussian fit; and the yellow line is the cumulative distribution function (cdf, on the right axis). The vertical lines denote the standard deviation, 95%, 99%, 99.9%, and the full-width half maximum

(fwhm) of the background absorbance values. (b) Absorbance image at 1650 cm$^{-1}$ and cell segmentation using the 99% absorbance threshold. Red outlines show excluded cells that are cut off in the image, cyan outlines represent small objects that are filtered out, and magenta outlines show segmented cells used for protein mass measurement. (c) Histogram of protein mass measurements of cells shown in (b) using different absorbance threshold levels. The average protein mass varies from 169 pg to 159 pg. (d) Histogram of protein mass measurements normalized by segmentation area for different threshold levels. Changing threshold levels can significantly change the cell area but not significantly alter the protein mass measurement.

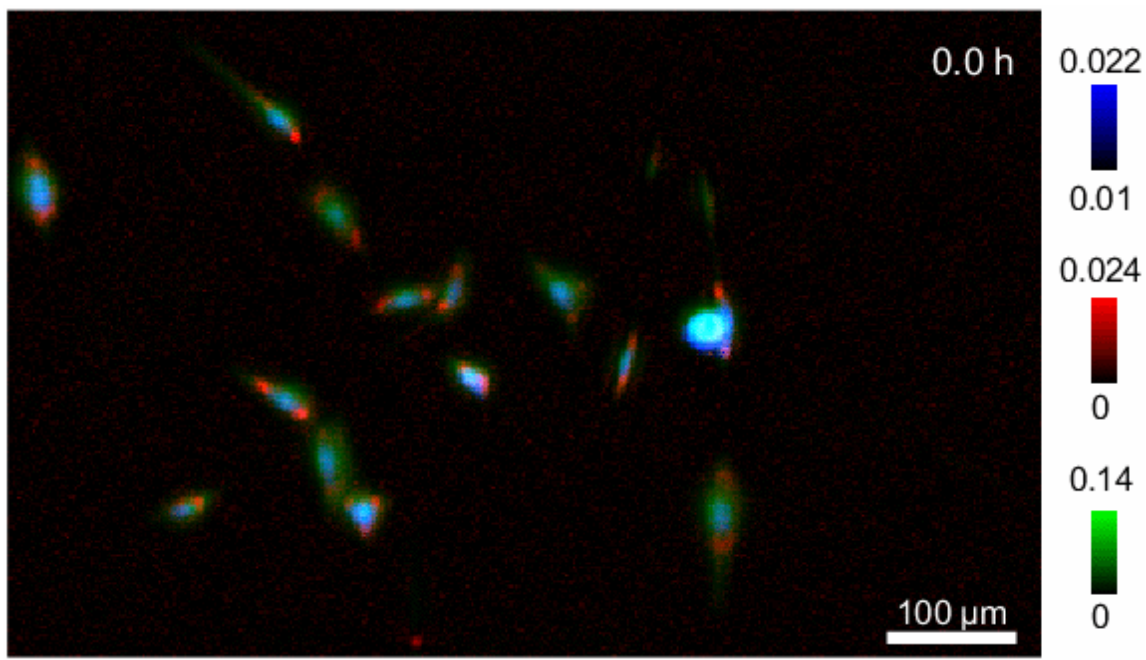

**SI Movie 1**. Live cell IR absorption movies of NIH 3T3 fibroblast cells. Green color represents absorbance associated with protein at 1656 cm$^{-1}$; red, for fatty acid at 1745 cm$^{-1}$; and blue, for nucleic acid at 1230 cm$^{-1}$.